\documentstyle[preprint,aps,pra]{revtex}
\begin{document}
\draft
\preprint{Proc. HCIS-X -- Berlin, 1997 (phys. stat. sol.(b), to be published)}
\title{
Coulomb-correlation effects on the non-linear \\
optical properties of realistic quantum wires
}
\author{Fausto Rossi\footnote{phone: +39 (59) 586072 --- fax: +39 (59) 367488 
-- E-mail: Rossi@UniMo.It} and Elisa Molinari}
\address{
Istituto Nazionale Fisica della Materia (INFM) and \\
Dipartimento di Fisica, Universit\`a di Modena \\ 
via Campi 213/A, I-41100 Modena, Italy 
}
\date{\today}
\maketitle
\begin{abstract}

We review recent results on the linear and non-linear optical response 
of realistic quantum-wire structures.
Our theoretical approach is based on a set of generalized semiconductor 
Bloch equations, and allows a full three-dimensional 
multisubband description of Coulomb correlation for any shape of the 
confinement profile, 
thus permitting a direct comparison with experiments for available 
state-of-the-art wire structures.

Our results show that electron-hole Coulomb correlation removes
the one-dimensional band-edge singularities from the absorption spectra, 
whose shape results to be heavily modified with respect to the ideal
free-particle case over the whole range of photoexcited carrier 
densities.

\end{abstract}
\clearpage
\narrowtext

\section{Introduction}

The dominant role played by Coulomb correlation in the optical response of
semiconductors and its dependence on dimensionality has now been long
recognized \cite{Books}. 
More recently, increasing interest has been devoted to one-dimensional (1D) 
structures \cite{Review_Wires}, prompted by 
promising advances in quantum-wire fabrication 
and application, e.g. in quantum-wire lasers.
The main goal of such effort in basic research as well as 
technological applications is to achieve structures with improved optical 
efficiency as compared to two-dimensional (2D) and three-dimensional (3D) ones. 

A common argument in favour of this effort is based on the  
well known van Hove divergence in the 1D joint density-of-states (DOS),  
which is expected to give rise to very sharp peaks in the optical spectra 
of 1D structures. Such prediction is however based on free-particle 
properties of ideal 1D systems and it ignores any  
disorder-induced and Coulomb-correlation effects.

As pointed out in the pioneering papers by Ogawa et al. \cite{Ogawa}, 
electron-hole correlation is 
expected to 
strongly influence the optical spectra of 1D systems. Their theoretical 
investigation ---based on a single-subband-model solution of the 1D 
Schr\"odinger equation in terms of a modified 1D Coulomb potential
\cite{Ogawa}--- shows that the inverse-square-root singularity in
the 1D DOS at the band edge is 
smoothed when electron-hole correlation is taken into account. 
The question is whether one can expect that the above
theoretical predictions, obtained for model 1D systems, also apply to 
the real
quantum wires made available by the present technology. Indeed, wires with
the best optical quality presently include structures obtained by epitaxial
growth on non-planar substrates (V-shaped wires)
\cite{Review_Wires,Kapon,Rinaldi}, or by cleaved-edge 
quantum well overgrowth (T-shaped wires) \cite{Wegscheider,Sakaki}. Owing to 
the shape of the confinement potential, these systems are far from an 
ideal 1D character. 
While quasi-1D confinement has been demonstrated for the lowest level
\cite{Review_Wires,Kapon,Rinaldi,Wegscheider,Sakaki}, 
excited states gradually approach a 2D-like behaviour. 
Moreover, in the available samples 
the subband separation is still relatively small, so that the 
coupling between different subbands may be important. 

From an experimental point of view, 
it is a matter of fact that, while 2D features are clearly 
observed in photoluminescence excitation spectra of quantum wells, 
so far no ``sharp" 1D features have been detected in the 
corresponding spectra of quantum wires. This is true despite the high 
quality of some of these structures, whose 1D character 
has been independently established by other methods 
\cite{Review_Wires,Kapon,Rinaldi,Wegscheider}.
However, the measured spectra are expected to be also strongly influenced 
by disorder-induced inhomogeneous broadening \cite{Vancouver} 
and, therefore, it has been sofar difficult to identify the 
role played by electron-hole correlation. 

From all these considerations, the following
questions need to be answered:
\begin{itemize}
\item[] 
{\it Are electron-hole correlation effects playing a dominant role also in
realistic quantum-wire structures ? If so, are these expected to hinder the 
possible advantages of the reduced dimensionality for relevant values of
temperature and carrier density ?}
\end{itemize}

\section{Theoretical approach}

To answer these questions, a full 3D approach for the analysis of Coulomb 
correlation in realistic quantum wires has been recently proposed 
\cite{PRL1,PRL2}.
This theoretical scheme, described in \onlinecite{PRL1}, is based 
on a generalization of the 
well-known semiconductor Bloch equations (SBE) \cite{Books} to the case of 
a multisubband wire.

More specifically, by denoting with $\{k_z\nu^{e/h}\}$ 
the free electron and hole states 
($k_z$ and $\nu^{e/h}$ being, respectively, the wavevector along
the wire direction, $z$, and the subband index corresponding to the
confinement potential in the $xy$ plane), we consider as kinetic variables the 
various distribution functions of electrons and holes 
$f^{e/h}_{k_z\nu}$ 
as well as the corresponding diagonal ($\nu^e = \nu^h = \nu$) 
interband polarizations 
$p^{ }_{k_z\nu}$.
This kinetic description is a generalization to 
1D systems of a standard approach for the study of bulk semiconductors 
\cite{Books} recently applied also to 
superlattice structures \cite{PRL_SL}.
Within our $k_z\nu$ representation, the SBE, describing the 
time evolution of the above kinetic variables, are written as 
\begin{eqnarray}
\label{SBE}
\frac{\partial}{\partial t} f^{e/h}_{\pm k_z\nu} 
&=& 
\frac{1}{i\hbar} \left({\cal U}^{ }_{k_z\nu} p^*_{k_z\nu} -
{\cal U}^*_{k_z\nu} p^{ }_{k_z\nu}\right) 
+ \frac{\partial}{\partial t} f^{e/h}_{\pm k_z\nu}\biggl|_{inco}  
\nonumber \\
\frac{\partial}{\partial t} p^{ }_{k_z\nu} 
&=& \frac{1}{i\hbar} 
\left({\cal E}^e_{k_z\nu} + {\cal E}^h_{-k_z\nu} \right)
p^{ }_{k_z\nu} 
+ \frac{1}{i\hbar} {\cal U}^{ }_{k_z\nu} 
\left(1 - f^e_{k_z\nu} - f^h_{-k_z\nu}\right) 
+ \frac{\partial}{\partial t} p^{ }_{k_z\nu}\biggl|_{inco} \ ,
\end{eqnarray}
where ${\cal U}^{ }_{k_z\nu}$ and ${\cal E}^{e/h}_{k_z\nu}$ 
are, respectively, the 
renormalized fields and subbands, whose explicit form involves the
full 3D Coulomb potential \cite{PRL1}.
The $\pm$ sign in Eq.~(\ref{SBE}) refers to electrons ($e$) and
holes ($h$), respectively, while the last terms on the rhs of Eq.~(\ref{SBE}) 
denote the contributions due to incoherent processes, e.g.
carrier-carrier and carrier-phonon scattering.

In this paper we focus on the quasi-equilibrium regime.
Therefore, Fermi-Dirac $f^{e/h}_{k_z\nu}$ are assumed
and the solution of the set of SBE (\ref{SBE}) simply reduces to 
the solution of the polarization equation. This is performed by means of a 
direct numerical evaluation of the stationary solutions, i.e. 
polarization eigenvalues and eigenvectors.
These two ingredients fully determine the absorption spectrum as well as
the exciton wavefunction in 3D real space. 
In particular, the electron-hole correlation function vs.
the relative free coordinate $z = z^e-z^h$  is given by:
$g(z) \propto \sum_{k_z k'_z \nu} p^*_{k_z\nu} p^{ }_{k'_z\nu} 
e^{i(k'_z-k_z) z}$.

The main ingredients entering our calculation are then the single-particle 
energies and wavefunctions, which in turn are numerically computed starting 
from the real shape of the 2D confinement 
potential deduced from TEM, as in Ref. \cite{Rinaldi}. 

\section{Numerical results} 

The above theoretical scheme has been applied to realistic V- and T-shaped wire
structures. In particular, here we show results for the 
GaAs/AlGaAs V-wires of Ref.~\cite{Rinaldi} and 
the GaAs/AlAs T-shaped wire of Ref.~\cite{Sakaki} 
(sample S2 [$d_1=d_2=53\AA$]). 

In order to better illustrate the role played by electron-hole correlation, 
in Fig.~1 we first show the
linear-absorption spectra obtained when taking into account the lowest wire 
subband only.
Here, results of our Coulomb-correlated (CC) approach are 
compared with those of the free-carrier (FC) model \cite{foot1}. 
For both V-shaped [Fig.~1(a)] and T-shaped [Fig.~1(b)] structures,
electron-hole correlation introduces two important effects: 
First, the excitonic peak arises below the onset of the continuum, with 
different values of binding energies (about $12$ and $16$\,meV, 
respectively) in good agreement with experiments \cite{Rinaldi}.
As discussed in \onlinecite{PRL2}, this difference is mainly ascribed to the 
different barrier height while the excitonic confinement is found to be 
shape (V vs. T) independent.
Second, we find a strong suppression of the 1D DOS singularity, in agreement 
with previous investigations based on simplified 1D models \cite{Ogawa}.

Let us now discuss the physical origin of the dramatic suppression of the
band-edge singularity in the CC absorption spectrum [solid lines in Fig.~1].
Since the optical absorption is proportional to the product of 
the electron-hole DOS and the oscillator strength (OS),
we have studied these two quantities separately. 
Figure 2(a) shows that the quantity which is mainly modified by CC is the OS. 
Here, the ratio between the CC and FC OS is plotted as a function of 
the excess energy with respect to the band edge (solid line). 
This ratio is always less than one and, in agreement with the 
results of 1D models \cite{Ogawa}, it goes to zero at the band edge. 
Such vanishing behaviour is found to dominate the 1D DOS singularity 
(dashed line)
and, as a result, the absorption spectrum exhibits 
a regular behaviour at the band edge [solid lines in Fig.~1].

Since the OS reflects the value of the correlation
function $g(z)$ for $z = 0$ \cite{Ogawa}, 
i.e. the probability of finding 
electron and hole at the same place, the vanishing behaviour of the OS in 
Fig.~2(a) seems to indicate a sort of electron-hole ``effective 
repulsion''. 
This is confirmed by a detailed analysis of the electron-hole correlation 
function,  $g(z)$,  reported in Fig.~2(b). 
Here  $g(z)$ (corresponding to the square of the exciton wavefunction in 
a 1D model) is plotted for three different values of the excess energy. 
We clearly see that the values of $g$ 
for $z = 0$ correspond to the values of the OS ratio at the same 
energies [solid line in Fig.~2(a)].
Moreover, we notice the presence of a true ``electron-hole correlation 
hole'', whose spatial extension strongly increases when approaching 
the band edge. 

The above analysis provides a positive answer to our first question: 
also for realistic quantum-wire structures electron-hole correlation leads 
to a strong suppression of the 1D band-edge singularity in the 
linear-absorption spectrum. Contrary to the 2D and 3D case, the Sommerfeld 
factor, i.e. the ratio between the CC and FC absorption, results to be less
than unity. 

Finally, in order to answer our second and more crucial question, we must 
consider that
most of the potential quantum-wire applications, i.e. 1D lasers and 
modulators, operate in strongly non-linear-response regimes
\cite{Review_Wires}.
In such conditions, the above linear-response analysis has to be 
generalized taking into account additional factors as:
(i) screening effects,
(ii) band renormalization,
(iii) space-phase filling.
We want to stress that all these effects are already accounted for in our
SBE (\ref{SBE}) \cite{PRL1}. 

Figure 3 shows the first quantitative analysis of non-linear absorption 
spectra of realistic V-shaped wire structures for different 
carrier densities at room temperature. 
In Fig.~3(a) we show as a reference the results obtained by including 
the lowest subband only. 
In the low-density limit (case A: $n = 10^4$ cm$^{-1}$) we clearly recognize 
the exciton peak. With increasing carrier density,
the strength of the excitonic absorption decreases due to phase-space 
filling and screening of the attractive electron-hole interaction, and 
moreover the band renormalization leads to a red-shift of the continuum.
Above the Mott density (here about $8*10^{5}cm^{-1}$), the exciton 
completely disappears.
At a density of $4*10^6$ cm$^{-1}$ (case D) the spectrum already exhibits a 
negative region corresponding to stimulated emission, i.e. gain regime. 
As desired, the well pronounced gain spectrum extends over a limited energy 
region (smaller than the thermal energy); However, its shape
results to differ considerably from the ideal FC one. The FC result is 
plotted in the same figure and marked with diamonds; note that it has been 
shifted in energy to align the onset of the absorption,
to allow a better comparison of the line-shapes \cite{foot2}.
The typical shape of the band-edge singularity in the ideal FC gain spectrum 
results to be strongly modified by electron-hole correlation. 
Also at this relatively high carrier density the OS corresponding to the CC
model goes to zero at the band edge as for the low-density limit 
[Fig.~2(a)]. As a consequence, the FC peak is strongly suppressed and only 
its high-energy part survives. The overall effect is a broader and less 
pronounced gain region.

Finally, Fig.~3(b) shows the non-linear spectra corresponding to the 
realistic case of a 12-subband V-shaped wire.
In comparison with the single-subband case [Fig3(a)], 
the multisubband nature is found to play an important role 
in modifying the typical shape of the gain spectra, which for both CC and FC 
models result to extend over a range much larger 
than that of the single-subband case for the present wire geometry [Fig.~3(a)].
In addition, the Coulomb-induced suppression of the single-subband 
singularities, here also due to intersubband-coupling effects, tends to 
further reduce the residual structures in the gain profile.
Thus, even in the ideal case of a quantum wire with negligible 
disorder- and scattering-induced broadening, our analysis indicates
that, for the typical structure considered, 
the shape of the absorption spectra over the whole density range 
differs considerably from the relatively sharp FC spectrum of Fig.~1.

\section{Conclusions}

We have presented a theoretical analysis of the linear and
non-linear optical properties of realistic quantum wires. 
Our approach is based on a numerical solution of the semiconductor Bloch
equations describing the multisubband 1D system.
We have applied such approach to typical T-  and V-shaped
structures, whose parameters reflect the current state-of-the-art in the
quantum-wire fabrication.

Our results allow us to reconsider the perspectives of quantum-wire 
physics and technology.
In particular, comparing the non-linear absorption spectra of Fig.~3(a) and
(b), we see that the broad gain region in (b) is mainly ascribed to the 
multisubband nature or, more precisely, to the small intersubband splitting
compared to the single-subband gain range in (a).
This confirms that, in order to obtain sharp gain profiles,
one of the basic steps in quantum-wire technology is to produce structures 
with increased subband splitting. 
However, the disorder-induced inhomogeneous broadening, not considered 
here, is known to increase significantly the spectral broadening 
\cite{Vancouver} and this 
effect is expected to increase with increasing subband splitting. 
Therefore, extremely-high-quality structures (e.g. single-monolayer 
control) seem to be the only possible candidates for successful quantum-wire 
applications.

\section*{Acknowledgments}

We thank Roberto Cingolani and Guido Goldoni for stimulating and fruitful 
discussions. 
This work was supported in part by the EC Commission through the 
Network ``Ultrafast''.
\clearpage
\begin{figure}
\caption{
Comparison between the absorption spectra obtained by including electron-hole 
Coulomb correlation (CC model, solid line) or by assuming free carriers
(FC model, dashed line)
for the case of: (a) the GaAs/AlGaAs V-wires of 
Ref.~\protect\cite{Rinaldi}\protect; 
(b) the GaAs/AlAs T-shaped wire of Ref.~\protect\cite{Sakaki}\protect 
(see text.) 
All spectra were computed assuming an energy broadening of $2$\,meV.
}
\end{figure}
\begin{figure}
\caption{
(a) Oscillator-strength ratio between CC and FC spectra (solid line) 
and electron-hole DOS (dashed 
line) as a function of the excess energy. (b) Electron-hole correlation 
function $g(z)$ vs. relative distance $z$ for three different values of the 
excess energy, identified by the corresponding symbols in (a). 
Note that $g(z=0)$ gives directly the oscillator strength 
for the corresponding excess energy (see text).
}
\end{figure}
\begin{figure}
\caption{
Non-linear absorption spectra of the V-shaped wire of Fig.~1(a) 
at room temperature 
for different carrier densities: 
A = $10^{4}$ cm$^{-1}$;
B = $5*10^{5}$ cm$^{-1}$;
C = $10^{6}$ cm$^{-1}$;
D = $4*10^{6}$ cm$^{-1}$;
E = $2*10^{7}$ cm$^{-1}$.
(a) single-subband case;
(b) realistic 12-subband case.
For the highest densities [case D in (a) and E in (b)]  
the corresponding FC result is also shown (line 
marked with diamonds). For a better comparison of the lineshapes, 
the FC band edge has been red-shifted to align with the corresponding CC one.
}
\end{figure}

\begin{references}
\bibitem{foot1} 
All the 
spectra were computed assuming a phenomenological energy broadening of 
2 meV. This small broadening (as compared to that of realistic wire 
structures \protect\cite{Vancouver}\protect) 
allows to better identify the role 
played by electron-hole correlation.
\end{references}
\end{document}